\documentclass[useAMS,usenatbib]{mn2e}
\usepackage{graphicx}
\usepackage{float}
\usepackage{caption}
\usepackage{subcaption}
\usepackage[authoryear]{natbib}
\usepackage{color}

\title[3D MHD study of the SN\,1006]{3D MHD simulation of polarized
emission in SN\,1006}
\author[E. M. Schneiter]{E. M. Schneiter$^{1,2,3}$\thanks{Email: mschneiter@gmail.com},
  P. F. Vel\'azquez$^{4}$,
  E. M. Reynoso$^{5,6}$, A. Esquivel$^{4}$, \and F. De
  Colle$^{4}$\\ $^{1}$Instituto de Astronom\'\i a Te\'orica y
  Experimental, Universidad Nacional de C\'ordoba, C\'ordoba,
  Argentina\\ $^{2}$Departamento de Materiales y Tecnologia, UNC,
  C\'ordoba, Argentina \\ $^{3}$ Deparment of Astronomy \& Oskar Klein
  Centre, Albanova, Stockholm University, SE-106 91 Stockholm, Sweden
  \\ $^{4}$Instituto de Ciencias Nucleares, Universidad Nacional de
  Mexico, Mexico D. F. Mexico\\ $^{5}$Instituto de Astronom\'\i a y
  F\'\i sica del Espacio, Suc. 28, CP: 1428, Buenos Aires, Argentina\\
  $^{6}$Physics Department, Faculty of Exact and Natural Sciences, University 
  of Buenos Aires, Argentina}
\begin{document}

\date{Accepted  . Received }

\pagerange{\pageref{firstpage}--\pageref{lastpage}} \pubyear{2002}

\maketitle

\label{firstpage}

\begin{abstract}
We use three dimensional magnetohydrodynamic (MHD) simulations to
model the supernova remnant SN\,1006. From our numerical results, we
have carried out a polarization study, obtaining synthetic maps of the
polarized intensity, the Stokes parameter $Q$, and the
polar-referenced angle, which can be compared with observational
results. Synthetic maps were computed considering two possible
particle acceleration mechanisms: quasi-parallel and  
quasi-perpendicular. 
The comparison of synthetic maps of the Stokes parameter $Q$ maps with
observations proves to be a valuable tool to discern unambiguously
which mechanism is taking place in the remnant of SN\,1006, giving
strong support to the quasi-parallel model. 

\end{abstract}

\begin{keywords}
MHD--radiation mechanisms: general -- methods : numerical --
supernovae: individual: SN\,1006 --ISM: supernova remnants 
\end{keywords}

\section{Introduction}

\begin{figure}
  \centering
\includegraphics[width=0.5\textwidth]{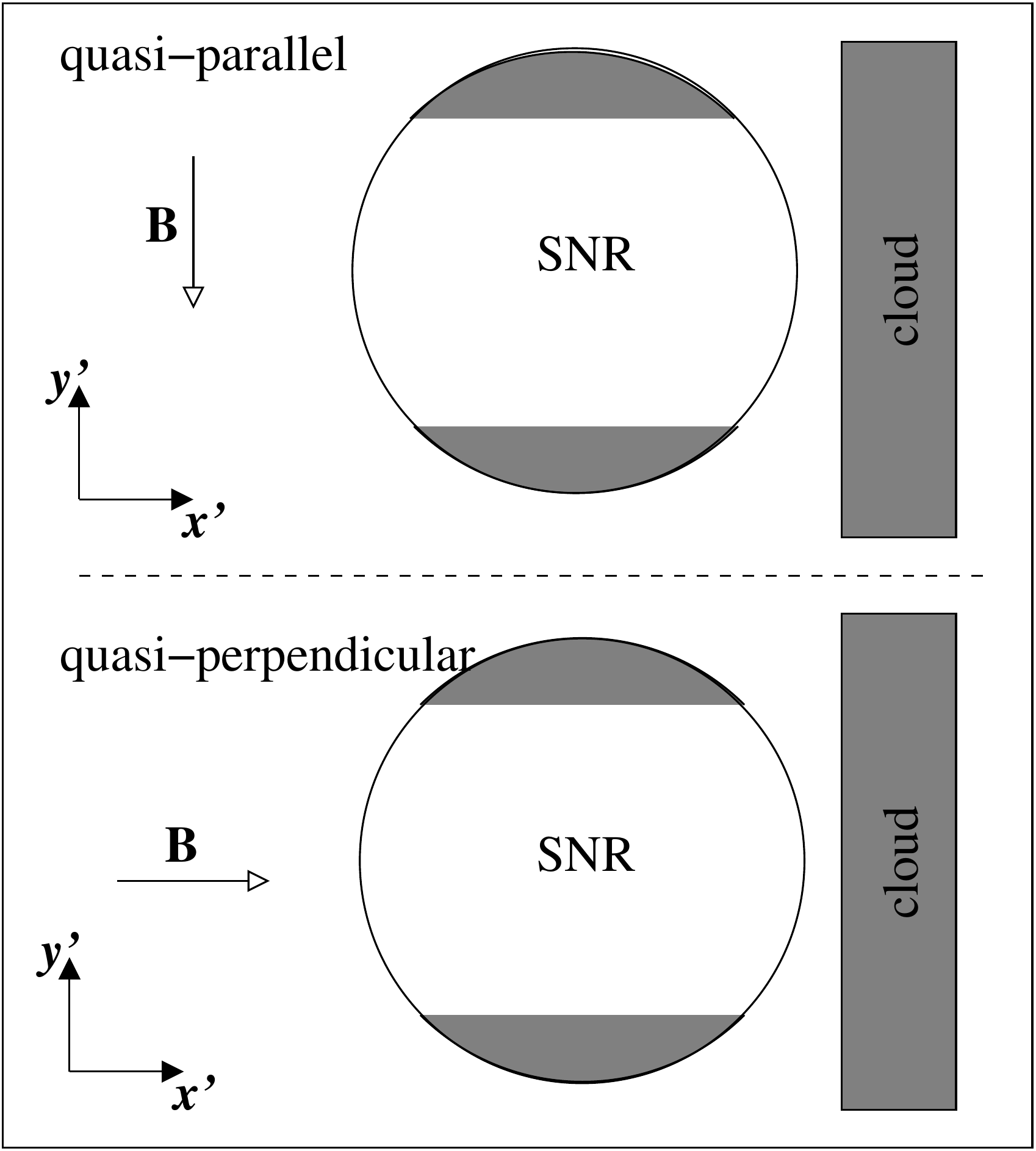}
\caption{Scheme of the numerical setup employed in our
  simulations for the quasi-parallel (top panel) and the
  quasi-perpendicular case (bottom panel). 
  In all the models a dense cloud was placed to the right.}
\label{scheme}
\end{figure}

Recently, the study of bilateral (also known as barrel-like) supernova
remnants (SNR) has gained great interest since they have proven to be a
useful tool when studying the configuration of the interstellar
magnetic field (ISMF) on scales of a few pc. As their name 
suggests, these type of SNRs display two characteristic bright
and opposite arcs. 
The morphology of this type of remnants has been
  explained with an ISMF, which is involved in the acceleration mechanism 
of relativistic particles.
However, there is still debate as to which is the most efficient
acceleration mechanism in these objects.
One interpretation is the equatorial belt model, in which the
orientation of the ISMF is quasi-perpendicular to the shock front
normal. The other interpretation is the polar cap model in which
they are quasi-parallel. 

The remnant of SN\,1006 is the archetype of the bilateral SNR group.
It has a diameter of 30\arcmin  (or 20 pc at a distance of 2.2 kpc, 
\citealt{winkler2003}) and, in radio-continuum and X-ray images, 
exhibits an incomplete shell with two bright arcs perpendicular to 
the Galactic plane \citep{reynolds1993,reynoso2013}.
SN\,1006 is accepted to be the result of a type Ia SN explosion
\citep{stephenson2002}. This
remnant is located at a high Galactic latitude, and therefore thought to
be evolving in an almost homogeneous interstellar medium (ISM). For
this reason, the radio-continuum morphology of the remnant must be
mostly affected by the characteristics of the surrounding ISMF.

Numerous theoretical and observational studies have been devoted to
analyze which mechanism is responsible for the morphology of this remnant, 
giving rise to two opposing interpretations: either the
quasi-perpendicular acceleration mechanism 
\citep[e.g.][]{fulbright1990,petruk2009,schneiter2010} or the
quasi-parallel mechanism
\citep[e.g.][]{volk2003,rothenflug2004,bocchino2011}.

Most of these works apply criteria based only on the brightness distribution 
of the synchrotron emission. Accordingly, their conclusions are strongly 
dependent on the orientation of the ISMF respect to the line of sight
({\it los}), and can not determine which particle acceleration mechanism
is actually taking place. 
A recent observational polarization study seems to favour the idea 
that the bright arcs of SN\,1006 can be explained by the polar cap
model, implying an ISMF parallel to the Galactic plane \citep{reynoso2013}. 

The present work is an effort to clarify which process or mechanism is
more suitable to explain the synchrotron emission of the remnant of
SN\,1006.
For this purpose, we carried out 3D MHD simulations employing the same
scenario proposed in \citet{schneiter2010}.
From the numerical results, we performed a polarization analysis of the 
Stokes parameters $Q$ and $U$, as well as the
polar-referenced angle, thus facilitating the direct comparison with
observations.

The organization of this work is as follows: section \S
\ref{numerical} presents the initial setup for the MHD simulations,
section \S \ref{synthetic} explains how the synthetic maps for the
radio emission and Stokes $Q$ and $U$ parameters were calculated,
section \S \ref{results} introduces the results obtained which are
further discussed and summarized in section \S \ref{discussion}.

\section{The numerical model}\label{numerical}

\subsection{Initial setup}
The numerical simulations were carried out with the parallelized 3D
MHD code Mezcal \citep{decolle2006,decolle2008,decolle2012}. This code
solves the full set of ideal MHD equations in cartesian geometry
($x',y',z'$) with an adaptive mesh, and includes a cooling function to
account for radiative losses \citep{decolle2006}.  The computational
domain is a cube of $24$ pc per side
and it is discretized on a six level binary grid, with a maximum
resolution of $2.3\times 10^{-2}$ pc. All the outer boundaries were
set to an outflow condition (gradient free).

\subsection*{The SNR}

The physical parameters of the SNR setup are the same as those
presented in \citet{schneiter2010}, but to help the readability  
of the present paper, we reproduce them below.

A supernovae explosion is initialized by the deposition of $E_{o}=2.05\times 
10^{51}$ erg of energy in a radius of $R_0=0.65$ pc
located at the centre of the computational domain. The energy is
distributed such that $95\%$ of it is kinetic and the remaining $5\%$ is
thermal. 

The ejected mass was distributed in two parts: an inner homogeneous
sphere of radius $r_c$ containing $4/7$ths of the total mass
(M$_*=1.4$M$_\odot$) with a density $\rho_c$, and an outer shell
containing the remaining $3/7$ths of the mass following a power 
law ($\rho \propto r^{-7}$) as in \citet{jun1996a}.
The velocity has a linear profile with $r$, which reaches a
value of $v_0$ at $r=R_0$. The parameters $\rho_c$, $r_c$, and $v_0$
are functions of $E_0$, $M_*$, and $R_0$, and were computed
using Eqs.(1)-(3) of \citet{jun1996a}. 

\subsection*{The surrounding interstellar medium}

The surrounding ISM, where the SNR evolves, was simulated as an
homogeneous plasma with temperature T$_0=10^{4}$K and number
density $n_0=5\times 10^{-2} $cm$^{-3}$. At a distance of $\sim 8$pc 
in the $x'$-direction from the centre of the computational domain,
the density was increased by a factor of 3 \citep{acero2007}.  The
SNR will collide with this `wall', producing a bright H$\alpha$ emission
filament \citep{winkler2003}, which has X-ray and radio counterparts 
\citep{acero2007, schneiter2010}.

In order to simulate the synchroton emission of SN\,1006, based on
the results of \citet{schneiter2010} we have considered two
configurations for the magnetic field $\mathbf{B}$ depending on which
of the particle acceleration processes is assumed. For the
quasi-perpendicular case, $\mathbf{B}$ is along the $x'-$direction,
and it is along the $y'-$direction for the quasi-parallel case (the
bright arcs are in the $y'-$direction). In both cases, we have
considered a mean magnetic field magnitude of $2 \mu$G, while only for
the quasi-parallel case an additional $\nabla 
\mathbf{B}$, increasing toward the $-x' -$direction, was included
\citep[which corresponds to model GRAD2 of][]{bocchino2011}. Figure
\ref{scheme} shows an scheme of the initial numerical setup employed
in our simulations. 

\section{Synthetic total and polarized radio emission maps}\label{synthetic}


\begin{figure}
  \centering
  \includegraphics[width=0.486\textwidth]{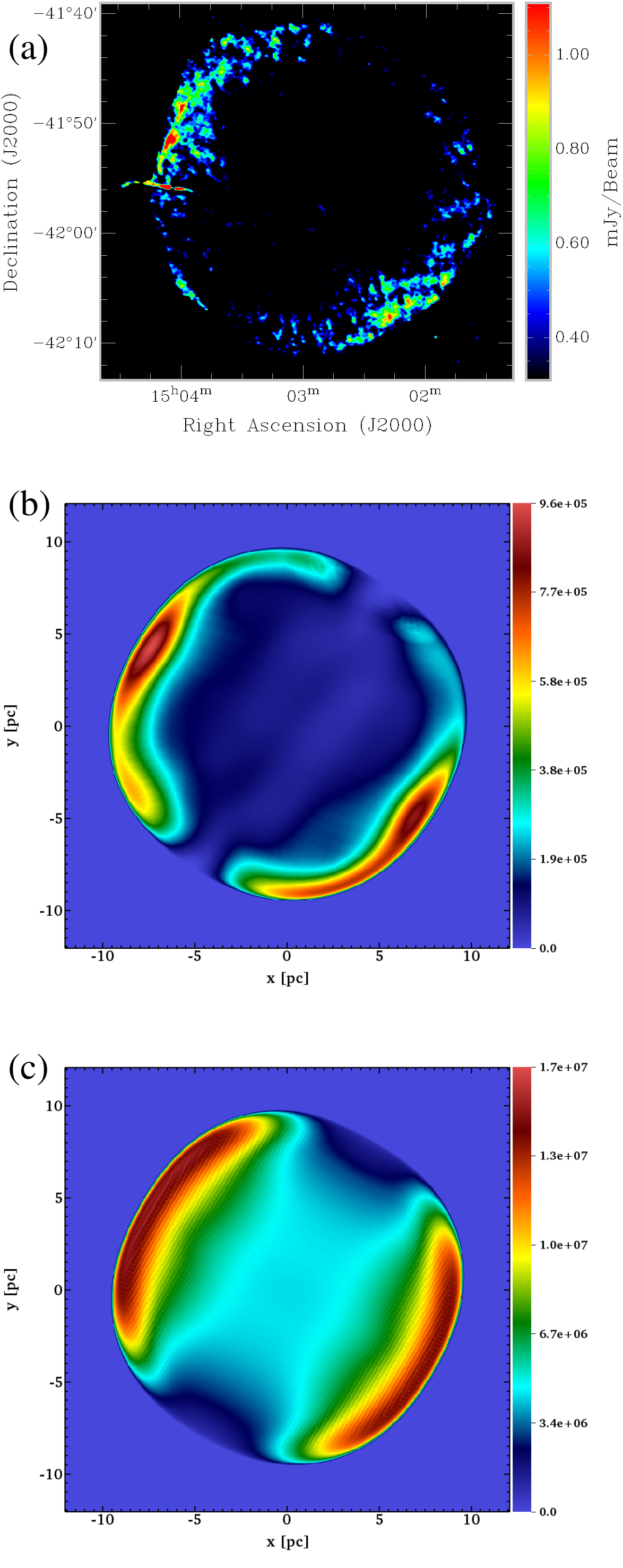}
    \caption{Synthetic maps of polarized intensity obtained for the
      quasi-parallel (panel b) and quasi-perpendicular (panel c) cases. 
The axes are in pc and the linear color scale is in arbitrary units. These maps 
were rotated $60$ degrees with the purpose of comparing them to a radio image 
at 1.4 GHz (panel a), displayed in equatorial coordinates. }
\label{fig:intpol}
\end{figure}


\subsection{Synchrotron emission maps}

The synchrotron emission is obtained as \citep{ginzburg65}:

\begin{equation}
i(\nu) \propto K B_{\perp}^{\alpha +1} \nu^{-\alpha}
\end{equation}
where $\nu$ is the observed frequency, $B_{\perp}$ is the component of
the magnetic field perpendicular to the {\it los}, and $\alpha$ is the
spectral index which was set to $0.6$ for this object. The coefficient $K\propto
\epsilon\ v_s^{-b}$ includes the dependence from the obliquity and the
shock velocity $v_s$ \citep{orlando2007}, where $\epsilon$ is either
proportional to $\sin^2{\Theta_{Bs}}$ for the quasi-perpendicular case or
$\cos^2{\Theta_{Bs}}$ for the quasi-parallel case. The angle $\Theta_{Bs}$ is
the angle between the shock normal and the post-shock magnetic field
\citep{fulbright1990}. As in \citet{orlando2007}, the exponent $b$ was
chosen to be $-1.5$, implying that stronger shocks are more efficient
injecting particles.

\subsection{Maps of the Stokes parameters}

To calculate the $Q$ and $U$ synthetic maps, the following expressions
for the Stokes parameters were employed \citep[see][]{clarke1989,jun1996b}

\begin{equation}
Q(\nu)=\int_{los} f_0  i(\nu) \cos\left[ 2\phi(s)\right] ds
\label{factorQ}
\end{equation}
\begin{equation}
U(\nu)=\int_{los} f_0  i(\nu) \sin\left[ 2\phi(s)\right] ds
\label{factorU}
\end{equation}
where $s$ increases along the {\it los}, $\phi(s)$ is the position
angle of the local magnetic field on the plane of the sky, and $f_0$
is the degree of linear polarization, which is function of the
spectral index $\alpha$:

\begin{equation}
f_0=\frac{\alpha +1}{\alpha + 5/3}
\end{equation}

The linearly polarized intensity is given by: 

\begin{equation}
I_P(\nu)= \sqrt{Q(\nu)^2+U(\nu)^2}
\label{ipol}
\end{equation}
and the polarization angle was computed as follows:

\begin{equation}
\chi=\frac{1}{2}\tan^{-1}(U/Q)
\label{chipol}
\end{equation}

\subsection{Observations}

\begin{figure}
        \centering
        \includegraphics[width=0.5\textwidth]{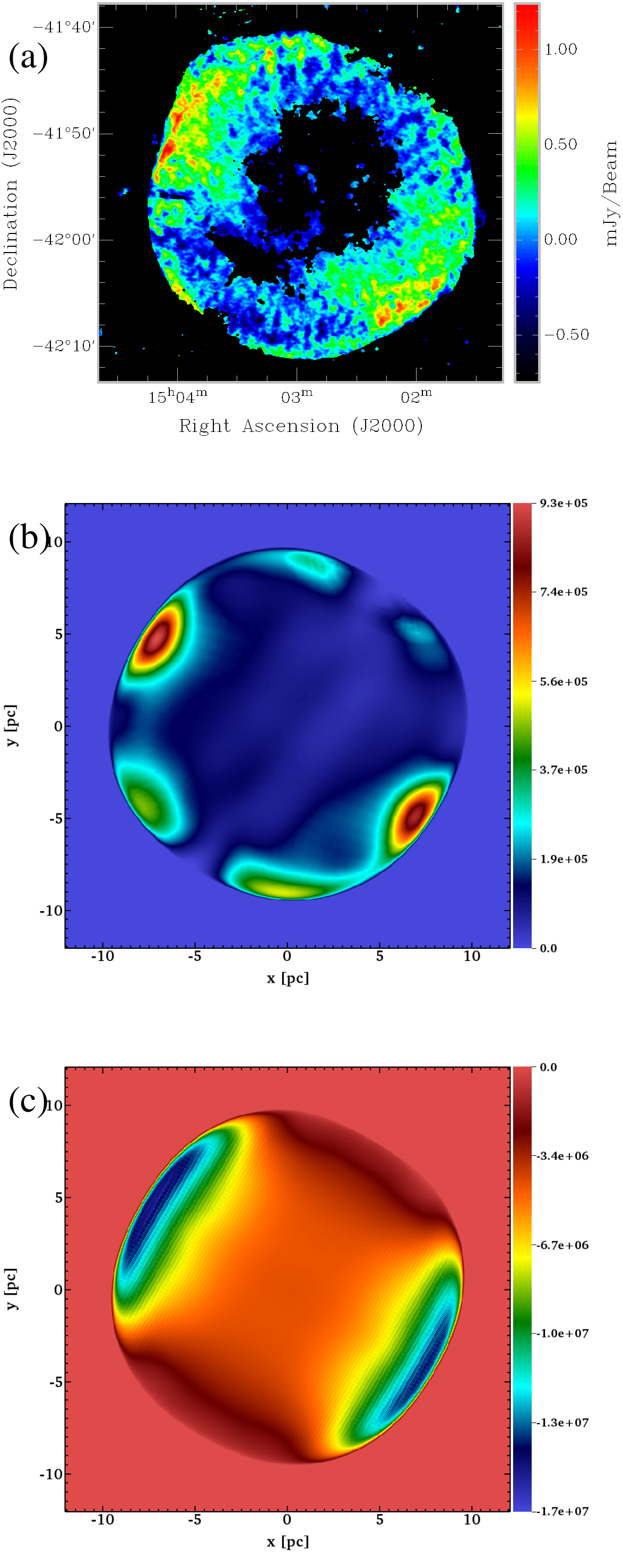}

        \caption{Comparison between the synthetic Stokes parameters Q
          for the quasi-parallel and the quasi-perpendicular cases
          (panels (b) and (c), respectively)
          with observations (panel a).}\label{fig:Q}
\end{figure}

In order to compare our simulations with SN\,1006 as observed in radio
wavelengths, we have made use of data obtained at 1.4 GHz with the 
Australia Telescope Compact Array (ATCA) and the Very Large Array (VLA), 
already published in \citet{reynoso2013}. These data contain information
on the polarization of the emission, thus we constructed a map of the
polarized intensity (Fig. \ref{fig:intpol}a) and another map 
with the Q parameter (Fig. \ref{fig:Q}a). The observations 
and data reduction process are described in \citet{reynoso2013}. The
images presented here are convolved with a beam of 15 arcsec. To remove
spurious features beyond the limits of SN\,1006, only pixels where the total 
intensity is above 75 mJy beam$^{-1}$ are retained.

\section{Results}\label{results}


\subsection{Synthetic polarization maps}
Synthetic polarized intensity maps were constructed from our
  numerical simulation results, using Eq.(\ref{ipol}).
Figure \ref{fig:intpol} displays a comparison of the observed
polarized emission (panel a), and the synthetic
maps  for the quasi-parallel (or polar cap
model, panel b) and quasi-perpendicular (or equatorial belt  model, panel c)
cases.
Before integrating along the {\it los} the
computational domain (the $x'y'z'-$ system) was tilted
$60\degr$ counterclockwise around the $z-$axis (the $los$), such that
the North points upwards and the East to the left (the plane of the
sky is the $xy$ plane). In addition, both models were rotated
$15\degr$ around the $y-$axis. The quasi-parallel model was further rotated
$15\degr$ in the $x-$axis.  Both synthetic maps exhibit two bright
arcs in the NE and SW direction, resembling the observations. The magnetic 
field gradient along the $x-$direction introduced in the simulations 
produces a SW-NE asymmetry in the polar cap model, in agreement with 
\citet{bocchino2011} and with observations \citep[see][and Figure 
\ref{fig:intpol} in this paper]{reynoso2013}.
As expected, the equatorial belt model shows greater
intensity due to a higher efficiency in diffusive shock 
acceleration \citep{jokipii1987,rothenflug2004, cassam2008}.

\subsection{Stokes parameter $Q$}

In order to compare the simulated maps of the Stokes parameter $Q$ with the
observational one, we had to take into account that the
  observation of radio polarization from the SNR actually gives the
  position angle of the electric field. Hence, we had to
  follow the inverse process to the observational one, that is, the
  magnetic field was rotated clockwise 90\degr, emulating the
  perpendicular electric field, and then we performed the
  inverse Faraday correction. To clarify, we have replaced the argument
  $\phi(s)$ in the trigonometric functions of Eqs. (\ref{factorQ}) and
  (\ref{factorU}) by $\phi(s)-\pi/2+\Delta\chi_{\mathrm F}$, where
  $\Delta\chi_{\mathrm F}$ is the Faraday rotation correction, 
  given by:
\begin{equation}
\Delta\chi_{\mathrm F}=\frac{\mathrm{RM}}{[\mathrm{rad\ m^{-2}}]}
\biggl(\frac{\lambda}{[\mathrm m]}\biggr)^2
\label{rotfaraday}
\end{equation}
being $\mathrm{RM}$ the rotation measure and $\lambda$ the observed wavelength.
For the case of the SN\,1006, \citet{reynoso2013} obtained an average
$\mathrm{RM}$ of 12\ rad m$^{-2}$ from their polarization study at
$\lambda=0.21$m.
With these values, Eq.(\ref{rotfaraday}) gives a correction angle due
to Faraday rotation of 0.52 rad or 30.3\degr.

Figure \ref{fig:Q} shows the resulting synthetic maps of the $Q$ parameter 
(panels (b) and (c) for the quasi-parallel and the quasi-perpendicular cases, 
respectively),  which were compared to the observed map (panel a). 

As mentioned above, the question about which process of acceleration
of relativistic particles is responsible for the synchrotron emission
in SNRs
\citep[see][]{fulbright1990,petruk2009,schneiter2010,volk2003,rothenflug2004,bocchino2011}
is an ongoing discussion in the literature. Our simulated polarized
intensity maps are not conclusive on which of the two competing
processes better explains the case of SN\,1006.
However, the maps of the Stokes parameter $Q$ are quite different between
the two models, thus removing the discrepancy. Hence,
the Stokes parameters turn out to be an important tool to compare
numerical models with observations and determine which of the
acceleration processes is more suitable in each scenario.

\subsection{Polar-referenced angles}

Following \citet{reynoso2013}, we created maps of the angular
difference for the position angle $\chi$  with respect to the local
radial direction $\hat r$, i.e. we obtained maps with the 
distribution of the polar-referenced angle $\chi_r$, which is given by:
\begin{equation}
 \chi_r=\cos^{-1}(\hat r . {\hat b}_{\perp}) 
\label{chir}
\end{equation}
where ${\hat b}_{\perp}$ is the direction of the magnetic field on the
plane of the sky. In this representation, $\chi_r = 0\degr$ and 
$\chi_r = \pm 90\degr$ correspond to the radial and tangential directions
respectively.  
These maps allow us to analyze two aspects of the magnetic field: the 
orientation of the post-shocked magnetic field and the orientation of the 
unshocked magnetic field of the ambient ISM.  We expect the magnetic fields 
in young SNRs to be radially aligned due to the development of hydrostatic 
instabilities \citep{jun1996b}.  So, the polar-referenced angle, $\chi_r$, 
should have a value close to zero on the edge of the shell.

\begin{figure}
  \centering
   \includegraphics[width=0.5\textwidth]{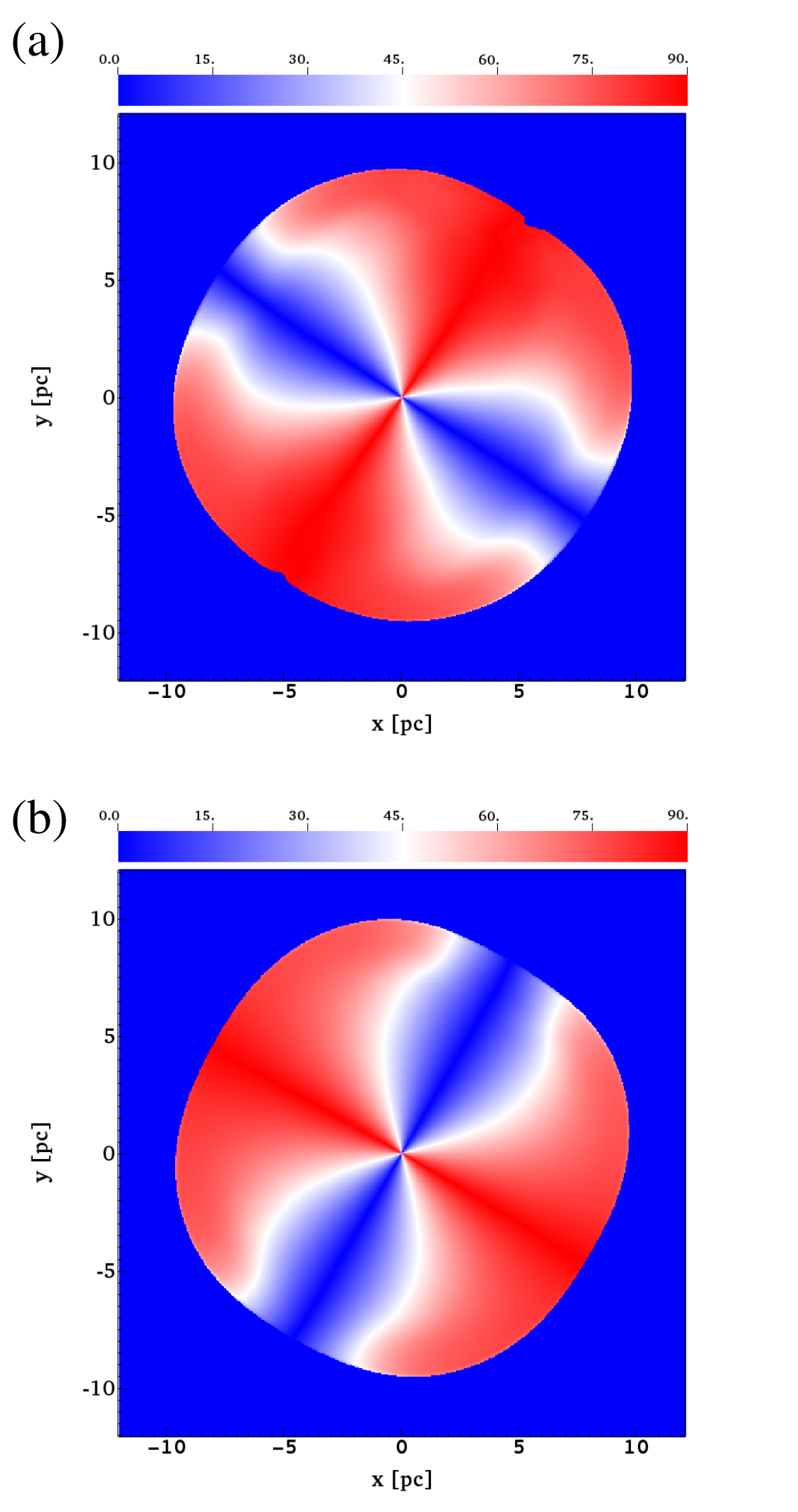}
    \caption{Synthetic polar-referenced angle maps obtained for the
      quasi-parallel (panel a) and quasi-perpendicular (panel b) cases. These maps were
      rotated $60$ degrees with the purpose of comparing them to the
      observations. The axes are in pc and the linear color scale is
      given in degrees.}
\label{refangles}
\end{figure}

Figure \ref{refangles} shows maps of the distribution of $\chi_r$,
obtained for both quasi-parallel and quasi-perpendicular cases. These
maps show a very small region on the SNR periphery with radial
magnetic field orientation, and $\chi_r$ takes values that run from
0\degr (blue color), to 90\degr (red color).  For both cases, the
regions with $\chi_r\simeq 0$ are nearly aligned with the direction of
the unperturbed magnetic field as imposed in each model.  This
`radial' region is tilted clockwise $\sim30\degr$ respect to the $\hat
y$ or `North' direction (see panels (b) of Figure \ref{refangles}) for
the quasi-perpendicular case, and $\sim 60\degr$ counterclockwise for
the quasi-parallel case (see panel (a) of Figure \ref{refangles}).
The quasi-parallel model gives a very good agreement (at large scales)
with the observations \citep[see Figures 7(a) and (b)
  in][]{reynoso2013}, suggesting that the direction of the ambient
magnetic field proposed in this model must be representative of the
actual orientation.  We must note however, that the polar-referenced
angles show some spread at small scales. This is most likely due to
inhomogeneities of the magnetic field at such scales, which is not
accounted for in our simulations.

\section{Discussion and Conclusions}\label{discussion}


We have utilized 3D MHD models to simulate the polarized emission of 
SN\,1006.  The 3D simulation{\bf s} allow us to explore configurations that 
were impossible with the axisymmetric code used in \citet{schneiter2010}.  
In particular it is possible to test different orientations of the magnetic 
field with respect to the plane of the sky and avoids some artifacts that 
arise when generating synthetic maps.

SN\,1006 is a member of a population of `barrel-like' or `bilateral' 
SNRs. Previous studies
\citep{reynolds1993,petruk2009,schneiter2010,bocchino2011,reynoso2013}
have shown that the emission geometry of bilateral SNRs is correlated 
with the orientation of the ambient interstellar magnetic field. 
Different orientations of the magnetic field can be inferred depending upon 
which particle acceleration mechanism is assumed to be taking place at the 
SNR shock front, either the quasi-parallel or the quasi-perpendicular model. 

As mentioned above, there is an ongoing debate as to which of these
two particle acceleration mechanisms is
responsible for the observed syncrotron emission in SNRs.  While
observational works seem to support the quasi-parallel model (polar
cap scenario), theoretical studies are not conclusive. A probable
reason for this  discrepancy is that most of the theoretical works 
base their conclusions on criteria which only take into account the
brightness distribution of the synchrotron emission. 

In the present work, we explore if a theoretical polarization study
can be a useful tool to shed some light on this question.
We synthesized maps of the polarized emission that include Stokes
$Q$ and $U$, and the distribution of the polar-referenced angle $\chi_r$.
In our analysis, we considered two scenarios: (1) {\bf an} equatorial 
belt model where the bright rims of SN\,1006 are formed
from particle acceleration that occurs when the shock passes through 
interstellar magnetic fields that are perpendicular to the shock normal 
(quasi-perpendicular); 
(2) a polar cap model where the bright rims are formed from
acceleration that occurs when the shock passes through magnetic fields that are 
parallel to the shock normal (quasi-parallel).

The resulting polarized intensity maps alone are not sufficient to
decide which mechanism is the adequate to explain the radio
morphology of SN\,1006 since both scenarios, quasi-parallel and
quasi-perpendicular, yield similar results.
On the contrary, the synthetic $Q$ maps for the quasi-parallel and 
quasi-perpendicular cases differ much from each other. The
quasi-perpendicular $Q$ map displays regions with negative
values resembling the bright rims observed in the polarized
intensity maps. This result is in disagreement with the observations,
which mainly display positive values \citep[Fig. \ref{fig:Q}; see also][]
{reynolds1993}. In addition, our simulated $Q$ map for the
quasi-parallel case coincides with observations in that the maxima are 
spatially coincident with the bright arcs of SN\,1006, as obtained
in our synthetic $Q$ map for the quasi-parallel case. Finally,
the synthetic polar-referenced angle map for the
quasi-parallel model is in very good agreement with the results
reported by \citet{reynoso2013}.
  
Our results agree with \citet{bocchino2011} in that the ISM magnetic
field must be parallel to the galactic plane, tilted with respect to
the {\it los}, and with a gradient that explains the assymetry in the
emission between the arcs. The difference between their result and
ours lies in the degree of inclination with respect to the {\it los}.
For the parameter $Q$ to be comparable to the observations, we
employed an inclination of 75\degr with respect to the $los$ (15\degr
respect to the plane of the sky), which is somewhat larger than the
best fit ($38\degr \pm 4\degr$) reported in
\citet{bocchino2011}. 

We calculated the  polar-referenced angle ($\chi_r$)
  distribution and found a good agreement with the observations of
  SN\,1006 of \citet{reynoso2013}. The maps of observed $\chi_r$ show
  small scale structure that is absent on our models due to the well
  ordered field imposed. This technique might prove helpful
  for future studies that include the inhomogeneity of the media or of
  the magnetic field structure.

In summary, our polarization results provide a strong support to the
quasi parallel model, which is in agreement with previous
observational \citep{reynoso2013} and theoretical works
\citep{bocchino2011}. More importantly, the Stokes parameter $Q$
proved to be a powerful tool to determine the particle
acceleration mechanism that best explains the observed morphology and
polarization seen in SN\,1006. We note that in a previous 2D MHD
numerical simulation, \citet{schneiter2010} concluded that the
quasi-perpendicular model was the most suitable to explain the
observations. Our paper shows the importance of using 3D
simulations, since limiting the analysis to what is possible with an
axisymmetric setup led to an opposite result.  
Only a 3D model can account for a magnetic field that is tilted with 
respect to the {\it los} and can provide a realistic description of the 
particle acceleration and synchrotron emission of an SNR.

\section*{Acknowledgments}

We thank the referee for his valuable comments which helped
  us to improve the original version of the manuscript.
EMS and EMR are Member of the Carrera del Investigador Cient\'\i fico,
CONICET (Argentina).
We acknowledge Enrique Palacios Boneta for maintaining the Linux
server where our simulations were carried out. 
EMS, PFV, AE, and FDC thank financial support from CONACyT (M\'exico) grants
167611 and 167625, CONACyT-CONICET grant CAR 190489 and
DGAPA-PAPIIT (UNAM) IG100214, IA101614, 101413, and 103315. EMR is supported by CONICET 
grant PIP 112-201207-00226.

\bibliographystyle{mn2e}

\end{document}